\documentclass[prl,aps,floats,superscriptaddress,floatfix,twocolumn,longbibliography]{revtex4-1}
\usepackage{amssymb,amsmath}
\usepackage{amsmath,amssymb}
\usepackage{enumitem,kantlipsum}
\usepackage{lipsum}
\usepackage{soul}
\usepackage{cancel}
\usepackage{gensymb}
\usepackage{graphicx}
\usepackage{psfrag}
\usepackage{color}
\usepackage{dcolumn}
\usepackage{bm}
\usepackage[normalem]{ulem}

\def\beq{\begin{equation}}
\def\eeq{\end{equation}}
\def\bea{\begin{eqnarray}}
\def\eea{\end{eqnarray}}

\usepackage{hyperref}
\hypersetup{
  colorlinks=true,
  citecolor=magenta,
  linkcolor=cyan,
}

\begin{document}

\title{Mobility-induced order in active XY spins on a substrate}
\author{Astik Haldar}\email{astik.haldar@gmail.com}
\affiliation{Theory Division, Saha Institute of Nuclear Physics, A CI of Homi Bhabha National Institute, 1/AF Bidhannagar, Calcutta 700064, West Bengal, India}
\author{Apurba Sarkar}\email{apurbaphysics391@gmail.com}
\affiliation{School of Mathematical \& Computational Sciences, Indian Association for the Cultivation of Science, Kolkata-700032, West Bengal, India}
\author{Swarnajit Chatterjee}\email{swarnajitchatterjee@gmail.com}
\affiliation{Center for Biophysics \& Department for Theoretical Physics, Saarland University, 66123 Saarbr\"ucken, Germany}
\author{Abhik Basu}\email{abhik.123@gmail.com, abhik.basu@saha.ac.in}
\affiliation{Theory Division, Saha Institute of Nuclear Physics, A CI of Homi Bhabha National Institute, 1/AF Bidhannagar, Calcutta 700064, West Bengal, India}

\begin{abstract}
 We elucidate that the nearly phase-ordered active XY spins in contact with a conserved, diffusing species on a substrate can be stable. 
For  wide-ranging model parameters,  it has stable uniform phases robust against noises. { These are distinguished by {\em generalized} quasi-long range (QLRO)} orientational order { logarithmically} stronger or weaker than { the well-known QLRO} in equilibrium,   together with miniscule (i.e., hyperuniform) or giant number fluctuations, respectively. {This illustrates}  a direct correspondence between the two. The scaling of {\em both} phase and density fluctuations in the stable phase-ordered states is {\em nonuniversal}: they depend on the nonlinear dynamical couplings.  For other parameters, it has no stable uniformly ordered phase.  Our model, a theory for {\em active spinners}, provides a minimal framework for wide-ranging systems, e.g., active superfluids on substrates, synchronization of oscillators, active carpets of cilia and bacterial flagella and  active membranes.
\end{abstract}

\maketitle



 { The subject of stability and scaling in two-dimensional ($2d$) broken symmetry phase-ordered states  in rotationally invariant systems is central in wide-ranging {\em in-vivo} and {\em in-vitro} driven systems. }Prominent examples include driven $2d$ Bose systems~\cite{john-superfluid}, oscillator (i.e., rotating XY spins or rotors) synchronization phenomena in various systems~\cite{pikovsky-book,dorfler,strogatz,uriu1,uriu2,zhou,pagona,frasca,peruani,pagona,tirtha-oscii,batista}, ``active carpets'' of cilia or bacterial flagella, modeled as {\em active rotors} grafted on the carpet, which  can phase-order~\cite{ramin}, and take part in nutrient transport~\cite{lowen} or mucous in respiratory tract~\cite{volvox},  monolayers of spinning colloids~\cite{coll1,coll2}, and {\em in vitro} magnetic cilia carpets~\cite{mag-cilia}.  While it has been long believed that elasto-hydrodynamic interactions are crucial in synchronization of biological flagella~\cite{lauga1,yeomans1,saint1}, recently,  other mechanisms like  phase-dependent forces~\cite{ramin-bennet}  and ``cell-rocking''~\cite{geyer}  
 are shown to be crucial in the synchronization of swimming unicellular green alga {\em Chlamydomonas}.   While immobility is known to destroy synchronization in $2d$ isotropic systems~\cite{john-superfluid,dest1,dest2,dest3,dest4,dest5}, generic understanding of the impact of mobility and conservation laws on synchronization is still lacking. A comprehensive understanding of mobility-induced synchronization without direct hydrodynamic interactions in $2d$ systems is the central goal of this work.

\begin{widetext}
 
\begin{figure}[htb]
\includegraphics[width=\textwidth]{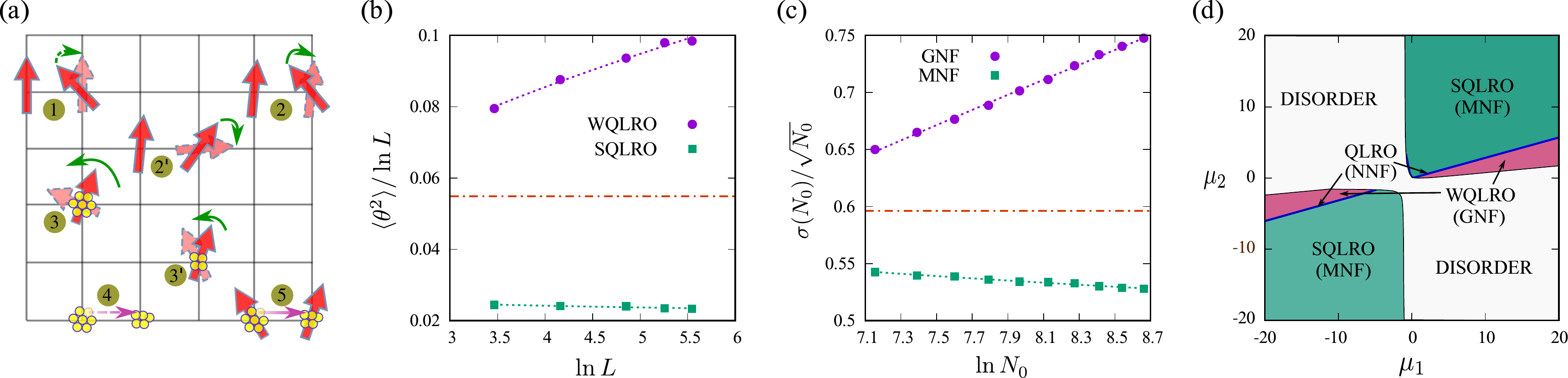}
\caption{ (a) Schematic diagram of the model showing the five update rules that define the model. The
phase changes by spin stiffness controlled local alignments reducing local phase differences (marked 1),  local phase difference (2 and 2$^\prime$) and local concentration (3 and 3$^\prime$) dependent active rotations. The dashed-outlined arrows represent the updated phase of the spins in response to the aforementioned factors. The mobile species moves by diffusion (4), and in response to local phase heterogeneities (5).
 Plots from agent-based simulations: (b) $\langle\theta^2\rangle/\ln L$ versus  $\ln L$ showing SQLRO and WQLRO, and  (c) $\sigma(N_0)/\sqrt{N_0}$ versus $\ln\,N_0$ showing MNF and GNF ($L=128$) with  $c_0=5,\,\xi=0.1$ {and}  (SQLRO/MNF): $g_1=1.0,\,g_2=0.02$,  and (WQLRO/GNF): $g_1=2.0,\,g_2=0.03$,  respectively.  The red broken horizontal lines in (b) and (c) correspond to the equilibrium results. The SQLRO-MNF and WQLRO-GNF correspondences are clearly established in the agent-based model (see text). (d) Schematic phase diagram in the $\mu_1-\mu_2$ plane  {in} the linearly stable hydrodynamic theory. SQLRO/MNF, WQLRO/GNF and disordered regions are marked in different colors. QLRO/NNF is the (blue) line demarcating  SQLRO/MNF and WQLRO/GNF regions. }  
\label{fig:modefigure}
\end{figure}

\end{widetext}

In this Letter, we explore  $2d$ phase-ordering or synchronization {\em without hydrodynamic interactions} in a $2d$ driven XY model interacting with a conserved, diffusive species on a substrate. We illuminate how the mobile species assists formation of stable phase-ordered states  qualitatively resembling quasi-long-range-order (QLRO).  We construct a conceptual agent-based lattice model, where XY spins (or rigid rods) are rigidly grafted uniformly on a $2d$ substrate of size $L\times L$, interacting with a diffusively mobile conserved species of mean concentration $c_0$ trapped to the substrate.  Inspired by bacterial quorum sensing and synchronization in response to complex and dynamic environments~\cite{quorum1,quorum2,quorum3}, we assume the microscopic phase and density dynamics 
 to depend on the local phase and density inhomogeneities in the agent-based model, 
as illustrated in Fig.~\ref{fig:modefigure}(a). Physically,  the ``activity'' of the model stems from   the propensity of the spins to rotate  in response to the local  concentration of the diffusing species and  the magnitude of the local phase difference.  Active rotation and the absence of any self-propulsion of the spins make this active model distinct from the celebrated ``moving XY''  Toner-Tu model~\cite{2d-polar}, or the microscopic Vicsek model~\cite{vicsek} for flocks.  We use it to characterize the scaling properties of the phase-ordered states. We further develop the hydrodynamic theory for the phase-ordered states, and validate the results  of the numerical investigations.  
Our theory provides a generic description of {\em active spinners, i.e., active spinning without directed walking} in the presence of conservation laws, a form of active matter that is largely unexplored.
Surprisingly, in this model 
the ordered states break conventional notions of universality, characterized by varieties of complex and heretofore unknown scaling behavior. 
 
   The update rules of density $c_\ell$ and phase $\theta_\ell$ at site $\ell$  must be invariant under translation and rotation of space, and rotating each spin by multiples of $2\pi$, and should break the clockwise-anticlockwise symmetry for (active) spin rotations. 
Phase  $\theta_\ell$ changes by local alignments due to spin stiffness,  and active rotations proportional to functions of the local phase differences and local concentrations.  For simplicity, we choose them to be  $g_1\cos (\sum_{\ell'}|\theta_\ell - \theta_{\ell'}|/4)$ and $g_2\Omega(c_\ell)$, respectively, where $\ell'$ is a nearest neighbor site of $\ell$; $g_1,\,g_2$ measure the relative amplitudes of these two active processes, and  can be of same or opposite signs, independently. Similar to the Vicsek model~\cite{vicsek}, we also add (small) noises to $\theta_\ell(t)$ during its time-update, representing nonequilibrium analogs of thermal noises in the system. Likewise, updates of $c_\ell$ take place via random hopping to the nearest neighbor sites with a lower density, and also via hopping controlled by a function of the  phase difference between the originating site $\ell$ and a randomly chosen nearest neighbor target site $v$. The latter is an analog of phase-dependent forces conceptualized in Ref.~\cite{ramin-bennet}. 
Again, this function is arbitrary, being constrained only by the above symmetries and should allow for generic rotation sense-dependence of the particle current due to the lack of the clockwise-anticlockwise symmetry. For simplicity, we take this function to be $\Omega^\prime (\langle c\rangle_v)\sin (\theta_\ell-\theta_v)$, where $\langle c\rangle_v$ is the mean density of sites $\ell$ and $v$. Functions $\Omega(c_\ell)$ and $\Omega^\prime(\langle c_v\rangle)$ are assumed to be quadratic functions of their arguments for simplicity. 
See associated long paper (ALP)~\cite{alp} for technical details.  Key elements of our model should be relevant  in wide variety of systems~\cite{bio1,bio2,bio3,bio4,bio5}. 




The most striking result from our model is that  mobility and number conservation together in this system can lead to {\em stable uniformly ordered phases} robust against finite noises and proliferation of topological defects (see supplemental movie MOV1~\cite{supp1}), whereas only short-range order (SRO) exists in the immobile limit (movie MOV2~\cite{supp1}). We characterize the steady states by the variance of the orientation fluctuations $\langle \theta^2\rangle \equiv \langle \sum_\ell (\theta_\ell -\overline \theta(t))^2\rangle/L^2$, $\overline \theta(t)\equiv \sum_\ell\theta_\ell(t)/L^2$  (this gives the mean spin rotation at time $t$),  and standard deviation $\sigma (N_0)\equiv \sqrt{\langle N^2\rangle - \langle N\rangle^2}$, where $N$ is the  total number of spins in an open square box of linear length $<L$, and $N_0\equiv \langle N\rangle$ is its average. Quite remarkably we systematically find that in this model, controlled by the model parameters, $\langle \theta^2\rangle$ grows either {\em faster} or {\em slower} than $\ln L$, together with  $\sigma (N_0)$ correspondingly growing {\em faster} or {\em slower} than $\sqrt{N_0}$, in contrast to the $\ln L$ (QLRO) and $\sqrt{N_0}$ (normal number fluctuations (NNF)) dependences, respectively, found in the equilibrium limit. Continuum, analytical approaches show (see below) $\langle \theta^2\rangle\sim (\ln L)^{\gamma_1},\;\sigma(N_0)\sim \sqrt{N_0/(\ln N_0)^{\gamma_2}}$, where $\gamma_1>0$ but $\gamma_2$ can be positive, zero or negative, are found to vary with $g_1,\,g_2$ and the parameters that define $\Omega^\prime(\langle c\rangle_v)$; $\gamma_1=1,\,\gamma_2=0$ correspond to QLRO phase fluctuations and NNF, respectively,  identical to the equilibrium case. 
Qualitatively speaking, although like QLRO, $\langle\theta^2\rangle$   grows with the system size $L$, it does so either  logarithmically {\em slower} than QLRO (i.e. stronger order than QLRO or ``SQLRO'', $0<\gamma_1<1$), or {\em faster} than QLRO (i.e. weaker order than QLRO or ``WQLRO'',\,$\gamma_1>1$), depending upon the model parameters.  In a  surprising correspondence with phase fluctuations, the  density fluctuations measured by $\sigma(N_0)$ are either { \em miniscule} or {\em hyperuniform} (MNF, $\gamma_2<0$), or {\em giant} (GNF, $\gamma_2>0$). Two representative plots from the simulations of the agent-based model in Fig.~\ref{fig:modefigure} (b) and (c) show the $ L$-dependence of $\langle\theta^2\rangle$ with SQLRO and WQLRO, and correspondingly, $N_0$-dependence of $\sigma(N_0)$ displaying MNF and GNF, respectively. These establish the SQLRO (WQLRO) and MNF (GNF) correspondence in the agent-based model~\footnote{The quantitative accuracy of these plots is restricted due to the limited range of $L$ used in the simulations}. This correspondence together with model parameter-dependent scaling exponents in the ordered states are  the unique features of our model. The  agent-based model also admits two kinds of disordered states; see ALP~\cite{alp} for details. 

We now outline the construction of the hydrodynamic theory  of the phase-ordered states that validates the above results, and quantitatively explains the indispensable role of the conserved density fluctuations to sustain order. Details can be found in ALP~\cite{alp}.

Due to the friction from the substrate, there is no momentum conservation, so the only
conserved variable on the surface is the number density $c({\bf x},t)$ of the conserved species  at $\bf x$ on the substrate at time $t$.  In addition, for nearly phase-ordered spins, the broken symmetry phase fluctuations $\theta({\bf x},t)$ about an arbitrary reference state are {\em slow, Goldstone} modes with relaxation rates diverging in the long wavelength limit, but the amplitude fluctuations relax {\em fast}. Therefore, $\theta ({\bf x},t)$, and  $c({\bf x},t)$ are the only hydrodynamic variables.  For a driven system, we must write down the equations of motion by appealing to the general symmetries of the phase-ordered state, the underlying microscopic dynamics, and conservation laws. Additional equilibrium requirements like detailed balance do not apply to our nonequilibrium system. 
 Retaining up to the lowest order symmetry-permitted nonlinear terms in $\theta$ and spatial gradients,  $\theta({\bf x},t)$ follows
 \begin{equation}
 \partial_t \theta = \kappa \nabla^2 \theta + \frac{\lambda}{2} ({\boldsymbol\nabla} \theta)^2 + \Omega(c) + f_\theta,\label{theta-full}
\end{equation}
where $\Omega (c)$ is a general function of $c$.  
Number density $c$ follows a conservation law with a current ${\bf J}_c$ given by
\begin{equation}
 {\bf J}_c=-D{\boldsymbol\nabla}c - \lambda_0\tilde \Omega(c){\boldsymbol\nabla}\theta,\label{c-curr}
\end{equation}
  truncating up to the lowest order in $\theta$ and spatial gradients; $\tilde \Omega(c)$ is yet another function of $c$.    
   While we have constructed (\ref{theta-full}) and (\ref{c-curr}) by writing down all the rotation-invariant leading order terms, each  term actually carries a simple physical interpretation. { The linear $\kappa\nabla^2\theta$ term in (\ref{theta-full}), and $D{\boldsymbol\nabla}c$ term in (\ref{c-curr}) are just the equilibrium spin relaxation and particle diffusion, respectively; $\kappa,\,D>0$ are the spin stiffness and diffusivity respectively.} The $\lambda$- and $\Omega(c)$-terms in (\ref{theta-full}) represent, respectively, active rotations of the spins in response to local gradients of $\theta$, and local spin concentration. The $\lambda_0$-term in (\ref{c-curr}) models particle currents in response to phases nonuniformities, and is reminiscent of ``phase-dependent forces''~\cite{ramin-bennet}. The two signs of $\lambda_0$ for a given $\tilde\Omega(c)$ in fact remind us of contractile and extensile active matters~\cite{sriram-RMP}. 
 Gaussian noises $f_\theta$ and $f_c$ are white and conserved noises, respectively, with zero mean, and variances
 \begin{eqnarray}
 &&\langle f_\theta({\bf x},t) f_\theta(0,0) \rangle = 2D_\theta \delta^2({\bf x}) \delta(t),\label{f-theta} \label{noise-theta}\\
 &&\langle f_c({\bf x},t) f_c(0,0) \rangle = 2D_c (-\nabla^2) \delta^2({\bf x}) \delta(t), \label{f-c}\label{noise-c}
\end{eqnarray}
consistent with    $\theta$ and $c$ being, respectively, a non-conserved  and a conserved variable.
Writing $c({\bf x},t)=c_0 + \delta c({\bf x},t),\,\langle\delta c\rangle=0$, we obtain 
\begin{align}
&\frac{\partial\theta}{\partial t}=\kappa\nabla^2\theta + \Omega_1\delta c+\frac{\lambda}{2} ({\boldsymbol\nabla}\theta)^2+ \Omega_2 (\delta c)^2 + f_\theta,\label{theta-full-eq}\\
&\frac{\partial\delta c}{\partial t}=\lambda_0\tilde\Omega_0 \nabla^2\theta + D\nabla^2\delta c + \lambda_0 \tilde \Omega_1 {\boldsymbol\nabla}\cdot(\delta c{\boldsymbol\nabla}\theta) + f_c,\label{c-full-eq}
\end{align}
where we have retained the most relevant nonlinear terms in fields and gradients. Parameters $\Omega_1\equiv \partial \Omega/\partial c|_{c=c_0},\,\Omega_2 \equiv \partial^2 \Omega/\partial^2 c|_{c=c_0},\,\tilde \Omega_0 \equiv \tilde \Omega|_{c=c_0},\,\tilde\Omega_1 \equiv \partial \tilde\Omega/\partial c|_{c=c_0}$.  Couplings $\lambda,\Omega_2$ and $\lambda_0\tilde\Omega_1$ have arbitrary signs.  Equations~(\ref{theta-full-eq}) and (\ref{c-full-eq}) together with  (\ref{noise-theta}) and (\ref{noise-c}) form the active spinner analog of the celebrated Toner-Tu theory for dry flocks~\cite{2d-polar}.
 Purely for symmetry reasons, Eqs.~(\ref{theta-full-eq}) and (\ref{c-full-eq}) describe the hydrodynamics of conserved, mobile active XY spins on a substrate with no other conserved quantity, and also immobile active XY spins interacting with an incompressible asymmetric binary fluid in its well-mixed phase on a substrate, where $c$ now is the binary fluid order parameter, and also chiral active hexatics on a substrate~\cite{maitra}. Equations~(\ref{theta-full-eq}) and (\ref{c-full-eq}), interestingly, describe the $2d$ Kardar-Parisi-Zhang (KPZ) equation~\cite{kpz,stanley} for surface growth or erosion (``sandblasting'')  of ``height'' $\theta({\bf x},t)$ with a conserved species on it, giving new insights on the suppression of the thresholdless instability of $2d$ KPZ surfaces,  and also an active fluid membrane with a finite tension but without momentum conservation that contains an active species in it~\cite{act-mem}. One-dimensional versions of (\ref{theta-full-eq}) and (\ref{c-full-eq}) are same as the hydrodynamic equations of a one-dimensional sedimenting crystal~\cite{rangan1}.

In the linearized limit of Eqs.~(\ref{theta-full-eq}) and (\ref{c-full-eq}), i.e., with $\lambda=\Omega_2=\tilde\Omega_1=0$,  small fluctuations around the linearly stable states ($\Omega_1\tilde\Omega_0\lambda_0>0$) travel non-dispersively with a wavespeed independent of wavevector $k$,  reminiscent of the traveling waves observed in synchronization in dense bacterial suspensions~\cite{bio2}. { The correlators in the linear theory are {\em exactly} calculated  conveniently in Fourier space. For instance, the}  phase and density  autocorrelation functions show QLRO and NNF, indistinguishable from the $2d$ equilibrium XY model and a non-critical equilibrium system with short-range interactions.  In particular, equal-time phase correlator 
\begin{equation}
  C^0_{\theta\theta}(k)\equiv \langle |\theta ({\bf k},t)|^2\rangle_0 \approx \frac{\overline D}{2 \Gamma k^2},\label{theta-k-corr}
\end{equation}
in the long wavevlength limit,
where $\overline D\equiv D_\theta + D_c$ and $\Gamma\equiv (\kappa+D)/2$; ``0'' refers to a linear theory result. This  in turn gives  for the variance 
 \begin{equation}
 \Delta_\theta^0\equiv \langle\theta^2 ({\bf x},t)\rangle_0 = \frac{\overline D}{4\pi \Gamma} \ln \left(\frac{L}{a_0}\right)\label{qlro-l}
\end{equation}
in $2d$;  $a_0$ is a small-scale cutoff.  Equation~(\ref{qlro-l}) corresponds to a logarithmically rough Edward-Wilkinson (EW)~\cite{EW,stanley} surface at $2d$. Furthermore
\begin{equation}
 C_{\theta\theta}^0(r)\equiv \langle\left[\theta({\bf x+r},t)-\theta({\bf x},t)\right]^2\rangle_0\approx \frac{\overline D}{2\pi\Gamma}\ln ( r/a_0) \label{theta-corr-l}
\end{equation}
for large $r$~\cite{chaikin}. Equations~(\ref{qlro-l}) and (\ref{theta-corr-l}) imply QLRO. The equal-time { density correlator} 
\begin{equation}
  C_{cc}^0(k)\equiv\langle |\delta c({\bf k},t)|^2\rangle_0 \approx \frac{\overline D}{2 \Gamma }\label{c-k-corr}
\end{equation}
is independent of $k$ for $k\rightarrow 0$. In real space, $C^0_{cc}(r)$ vanishes for $r\gg \zeta$, a microscopic length characterizing the short-range interactions. This further means 
$\sigma(N_0)$   scales with the mean $\langle N\rangle\equiv N_0$ as $\sqrt {N_0}$, giving NNF as expected in a non-critical equilibrium system with short-range interactions. If $\Omega_1\tilde\Omega_0\lambda_0<0$, small fluctuations about the uniform states are linearly unstable with growth rates proportional to $k$.

The Fluctuation-Dissipation-Theorem 
(FDT)~\cite{chaikin}   is broken  in the linearized theory.  This manifests in the non-vanishing cross-correlator $C_\times({\bf k})\equiv\langle\theta({\bf -k},t)\delta c({\bf k},t)\rangle$, { which is} a model parameter-dependent constant { in the limit of small $k$. }

It now behooves us to find whether nonlinear effects are relevant (in the RG or renormalization group sense), and  the scaling properties of any ordered states that are robust against finite noises. To study this, we perform one-loop perturbative RG analysis on Eqs.~(\ref{theta-full-eq}) and (\ref{c-full-eq}) at $2d$,  similar to the RG calculations on the KPZ equation~\cite{fns,stanley}, or the coupled Burgers-like equation for Magnetohydrodynamics~\cite{ab-jkb-sr,abfrey,abfrey1}.  It turns out, as   {discussed} below, that the nonlinearities  either introduce logarithmic modulations to the scaling of the linearly stable states, or to destroy those states altogether.

%
As usual, the RG is done by tracing over the short
wavelength Fourier modes of the fields~\cite{fns,stanley,halpin}, by expanding in the dimensionless coupling constant $g\sim \lambda^2\overline D/\Gamma^3$,  which is marginal in $2d$; see ALP~\cite{alp} for technical details. It predicts that not all but some of the linearly stable states are   robust against noises.  { This} is controlled by $\mu_1\equiv \Omega_2/\lambda,\,\mu_2\equiv \lambda_0\tilde \Omega_1/\lambda$, that are {\em marginal} (in the RG sense) at the one-loop order.  { We find} that for wide ranges of $\mu_1,\,\mu_2$, renormalized { scale-dependent} $g(k)= 1/[\Delta_1(\mu_1,\mu_2)\ln (\Lambda/k)]$ flows to zero {\em very slowly}, { where $\Delta_1>0$} as $k\rightarrow 0$ at $2d$, 
under successive applications of the RG procedure,  { for stable ordered phases}; { $\Lambda=2\pi/a_0$ is an upper wavevector cutoff}.  This gives, { as obtained} from the RG flow equations,  that renormalized, scale-dependent $\Gamma(k)=\Gamma [\ln(\Lambda/k)]^{\eta_2}$ and $\overline D(k)=\overline D[\ln(\Lambda/k)]^{\eta_1}$, both diverging logarithmically when $k\rightarrow 0$; $\eta_1(\mu_1,\mu_2),\,\eta_2(\mu_1,\mu_2)>0$ are constants related to $\Delta_1(>0)$. These log-divergences due to the ``slow'' vanishing of the coupling constant are reminiscent of the logarithmic anomalous elasticity in three-dimensional equilibrium smectics~\cite{pelc1,pelc2}.
 The resulting renormalized theory, owing to $g(k)\rightarrow 0$ as $k\rightarrow 0$, is {\em effectively linear}, albeit with renormalized parameters.  Indeed, straightforward calculations show that the renormalized correlation functions in the stable ordered states display surprising {\em logarithmic modulations of  the linear theory scaling}, giving order stronger or weaker than in the linear theory. 
  For instance, the renormalized phase correlator $C_{\theta\theta}^R(k)$ reads
 \begin{equation}
  C^R_{\theta\theta}(k) \approx \frac{\overline D}{2\Gamma  k^2[\ln (\Lambda/k)]^\eta},\label{theta-k-corr1}
\end{equation}
for $k\rightarrow 0$;  $\eta\equiv \eta_2-\eta_1$. Here and below,  $R$ refers to renormalized quantities. Exponent $\eta$ varies {\em continuously} with $\mu_1,\,\mu_2$ and can be positive or negative; detailed calculations show $\eta <1/3$ always~\cite{alp}.  For $\eta>(<)0$, $C^R_{\theta\theta}(k)\ll (\gg) C^0_{\theta\theta}(k)$ as $k\rightarrow 0$, demonstrating strong suppression (enhancement) of fluctuations in the long wavelength limit. 
Next, the renormalized variance $\Delta_\theta^R  \equiv\langle\theta^2 ({\bf x},t)\rangle_R $ now acquires a novel $L$-dependence:
\begin{equation}
 \Delta_\theta^R\approx\frac{\overline D}{4\pi\Gamma}\left[\ln\left(\frac{L}{a_0}\right)\right]^{1-\eta},\label{eta-intro}
\end{equation}
This means $\Delta_\theta^R$ can grow, respectively, {\em logarithmically slower} or {\em faster} with $\ln L$ than QLRO, representing order {\em stronger} or {\em weaker} than QLRO, named, respectively, SQLRO, or WQLRO. { This defines a hitherto unknown {\em generalized QLRO}, that evidently extends the well-known QLRO found in the equilibrium $2d$ XY model at low temperature to the realm of nonequilibrium.  We thus identify a {\em nonuniversal exponent} $\eta(\mu_1,\mu_2)$ expressed by Eq.~(\ref{eta-intro}) in our active XY model  that parametrizes the generalized QLRO. This exponent is universal in the equilibrium $2d$ XY model, i.e., Eq.~(\ref{eta-intro}) holds in equilibrium with  $\eta=0$, when QLRO is retrieved.}  Equation~(\ref{eta-intro}) also implies a surface logarithmically smoother or rougher than the $2d$ EW surface. Likewise,  the renormalized   correlator
 $C_{\theta\theta}^R(r)$, related to the inverse Fourier transform of $C_{\theta\theta}^R(k)$,  scales as
\begin{equation}
 C_{\theta\theta}^R(r)\equiv \langle[\theta({\bf x+r},t)-\theta({\bf x},t)]^2\rangle_R\approx\frac{\overline D}{2\pi\Gamma} [\ln(r/a_0)]^{1-\eta},\nonumber
\end{equation}
for large $r$.
Related to $C^R_{\theta\theta}(r)$, the renormalized   spin correlation function $C_{ZZ}^R(r)$  for large $r$  is
\begin{equation}
 C^R_{ZZ}(r)\equiv \langle \cos[\theta({\bf x+r},t)-\theta({\bf x},t)]\rangle_R \approx (r/a_0)^{-\tilde\gamma(r)} \label{czz}
\end{equation}
{where the {\em $r$-dependent exponent function} $\tilde \gamma(r)$  has a complex form: 
\begin{equation}
\tilde\gamma (r)\equiv \frac{\overline D}{4\pi\Gamma}[\ln(r/a_0)]^{-\eta}
\end{equation}
for large $r$,  {\em is nonuniversal}, varying continuously with $\mu_1,\mu_2$.  This generalizes the well-known parameter-dependent scaling of the spin-spin correlator in the equilibrium $2d$ XY model in its QLRO phase ($\eta=0$)~\cite{chaikin}} { and is yet another manifestation of the generalized QLRO displayed by our $2d$ active XY model}.
For $\eta>0(<0)$, clearly $C_{ZZ}^R(r)$ decays much slower (faster) for large $r$, giving SQLRO (WQLRO).

In contrast to its linear theory analog,  renormalized density correlator $C_{cc}^R(k)$  picks up a weak $k$-dependence:
\begin{equation}
C_{cc}^R(k)\approx \frac{\overline D}{2\Gamma}\left[\ln \left(\frac{\Lambda}{k}\right)\right]^{-\eta}
\end{equation}
in the hydrodynamic limit $k\rightarrow 0$. 
 Evidently, for $\eta>0$, the long wavelength density fluctuations are strongly suppressed vis-a-vis for $\eta =0$ (equilibrium limit result), implying MNF or hyperuniformity, an exotic state of matter~\cite{hyper,hyper-rev}  rarely encountered in ordered active matter~\cite{shelley}.  
 In contrast, if $\eta<0$ the density fluctuations are hugely enhanced when $k\rightarrow 0$. This is GNF, often encountered in orientationally ordered active fluids~\cite{sriram-RMP}, and also in equilibrium superfluids~\cite{chaikin}. 
Lastly, the equal-time  renormalized density autocorrelator $C_{cc}^R(r)$, the inverse Fourier transform of $C_{cc}^R(k)$,   is
\begin{equation}
 C^R_{cc}(r)\approx \frac{\overline D}{4\pi \Gamma }\frac{-\eta}{r^2 [\ln(r/a_0)]^{(1+\eta)}},\label{ren-c-corr}
\end{equation}
for large $r$, $r/a_0\gg 1$.
Thus for $\eta >(<) 0$, $C^R_{cc}(r)$ falls off relatively faster (slower)  for MNF (GNF). By using (\ref{ren-c-corr}) we re-express MNF (GNF) for $\eta>(<)0$: 
\begin{equation}
 \sigma(N_0)\propto \sqrt{ N_0/(\ln N_0)^\eta}<(>) \sqrt{N_0}, \label{num-fluc}
\end{equation}
in the renormalized theory.
For $\eta=0$, 
 (\ref{num-fluc}) reduces to that in an equilibrium system with  NNF having short-range interactions away from any critical point.

We thus show SQLRO (WQLRO) phase order is necessarily accompanied by MNF (GNF). 

Like $C_{cc}^R(k)$, renormalized cross-correlation function $C_\times^R(k)$ also picks up a weak $k$-dependence:  $C_\times^R(k)\propto \left[\ln(\frac{\Lambda}{k})\right]^{(1-3\eta)/2}\times {\cal O}(1)$ for small $k$. Thus FDT remains broken in the renormalized theory.

 The model shows breakdown of conventional dynamic scaling in the ordered phases: { The form of $\Gamma(k)$ in the renormalized theory shows that} the diffusive scaling of time $t$ with  $r$ in the linear theory changes to 
\begin{equation}
t\propto r^2 /[\ln (r/a_0)]^{(1-\eta)/2}\label{dyn-scaling}
\end{equation}
for large $r$. Thus the fluctuations relax   logarithmically faster than ordinary diffusion. Also, they relax faster with WQLRO/GNF ($\eta<0$) than SQLRO/MNF ($\eta>0$).  

For other choices of $\mu_1,\,\mu_2$, there are no stable ordered states due to the nonlinear effects. 
A phase diagram of the model in the $\mu_1-\mu_2$ plane over limited ranges of $\mu_1,\,\mu_2$ is shown in Fig.~\ref{fig:modefigure}(d).  Magnitudes of the model parameters, which parametrize scaling, should depend on $c_0$ and $a_0$ (a rough measure of the particle size) in experimental realizations of this model. Thus by changing $c_0$ and $a_0$ 
different regions of the phase space and hence different scaling behavior of the ordered states and the instabilities can be explored.

Heuristically, fluctuations of the conserved density $c$ leads to dynamically changing neighbors of each XY spin, which in turn creates an effective long-range interactions between phases at distant locations. This for appropriately chosen parameters can suppress the instability of the $2d$ KPZ equation, and sustain order. With  uniform density (i.e., $\delta c\equiv 0$),  Eq.~(\ref{theta-full-eq}) reduces to the KPZ equation with only SRO in $2d$~\cite{kpz,stanley,john-superfluid}, in agreement with our numerical studies   (movie MOV2~\cite{supp1}). 
 { We thus establish that a $2d$ active XY model in contact with a mobile, conserved species on a substrate can have stable phase-ordered states with complex scaling behavior marked by a heretofore generalized QLRO, concomitant with a novel phase-density fluctuation correspondences. The logarithmic modulation of the resulting scaling behavior is characterized by nonuniversal, model parameter-dependent nonuniversal scaling exponents.   These predictions are {\em exact} in the asymptotic long wavelength limit, as is shown in Ref.~\cite{alp} using
renormalization group arguments.} This study forms an archetypal example of nonequilibrium
phase-synchronization  without hydrodynamic interactions, { and is a rare example of how {\em interactions lead to continuously varying scaling exponents} in active systems.} 

{ From a phenomenological perspective, our results shed light on how a collection of beating cilia or flagella grafted on a carpet can phase-synchronize in the presence of a moving chemical on the carpet, a collective behavior of significance in living systems.} 
 We hope our theory will induce future experiments to measure $\langle \theta^2\rangle$ and $\sigma(N_0)$ in self-assembled spinners ~\cite{kotot}, rotor assemblies~\cite{robot-assem,shelley,rotor-transport}  and programmable magnetic cilia carpets~\cite{mag-cilia,coll1}.  

{\em Acknowledgement:-} AH and AB thank T. Banerjee and N. Sarkar for critical reading and helpful suggestions. AS thanks University Grants Commission (UGC), India and Indian Association for the Cultivation of Science, Kolkata for research fellowships. SC is financially supported by the German Research Foundation (DFG) within the Collaborative Research Center SFB 1027 and Indian Association for the Cultivation of Science, Kolkata. AS and SC also acknowledges Prof. R. Paul (IACS, Kolkata) for some invaluable suggestions and discussions and are also thankful to him for providing the computational resources.  AB thanks J. Toner for valuable  discussions in the early stages of this work, and comments concerning derivations of Eqs.~(\ref{ren-c-corr}) and (\ref{num-fluc}), and P. K. Mohanty, A. Maitra and  D. Levis for helpful comments, and the SERB, DST (India) for partial financial support through the MATRICS scheme [file no.: MTR/2020/000406]. AB has designed the problem, AH has contributed to the analytical part of the work, AS and SC have contributed equally to the numerical part of the work. All four authors jointly wrote the manuscript.

\bibliography{diffxyref}

\end{document}